\documentclass{ieeeaccess-fixed}
\usepackage{cite}
\usepackage{amsmath,amssymb,amsfonts}
\usepackage{algorithmic}
\usepackage{graphicx}
\usepackage{textcomp}

\usepackage{bm}
\makeatletter
\AtBeginDocument{\DeclareMathVersion{bold}
\SetSymbolFont{operators}{bold}{T1}{times}{b}{n}
\SetSymbolFont{NewLetters}{bold}{T1}{times}{b}{it}
\SetMathAlphabet{\mathrm}{bold}{T1}{times}{b}{n}
\SetMathAlphabet{\mathit}{bold}{T1}{times}{b}{it}
\SetMathAlphabet{\mathbf}{bold}{T1}{times}{b}{n}
\SetMathAlphabet{\mathtt}{bold}{OT1}{pcr}{b}{n}
\SetSymbolFont{symbols}{bold}{OMS}{cmsy}{b}{n}
\renewcommand\boldmath{\@nomath\boldmath\mathversion{bold}}}
\makeatother

\def\BibTeX{{\rm B\kern-.05em{\sc i\kern-.025em b}\kern-.08em
    T\kern-.1667em\lower.7ex\hbox{E}\kern-.125emX}}

\usepackage[]{CJK}

\usepackage{hyperref}
\hypersetup{
    colorlinks=true,
    citecolor=blue,
    linkcolor=blue,
    filecolor=magenta,
    urlcolor=blue,
    }

\usepackage{ragged2e}

\begin{document}
\history{Date of publication: December 15th, 2025}
\doi{\href{https://doi.org/10.48550/arXiv.2512.02679}{10.48550/arXiv.2512.02679}}

\title{Off-grid solar energy storage system with lithium iron phosphate (LFP) batteries in high mountains: a case report of Tianchi Lodge in Taiwan}
\author{\uppercase{Hsien-Ching Chung}\authorrefmark{\href{https://orcid.org/0000-0001-9364-8858}
{\includegraphics[height=6.5pt]{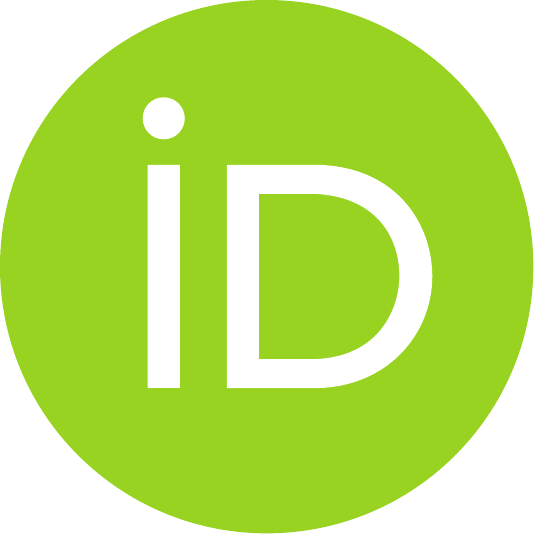}}1},
\IEEEmembership{Senior Member, IEEE}
}

\address[1]{Department of Research and Design, Super Double Power Technology Co., Ltd., Changhua City, Changhua County 500042, Taiwan (e-mail: hsienching.chung@gmail.com)}


\corresp{Corresponding author: Hsien-Ching Chung (e-mail: hsienching.chung@gmail.com).}

\begin{abstract}
Mountain huts are buildings located at high altitude, providing shelter and a place for hikers.
Energy supply on mountain huts remains an open issue.
Using renewable energies could be an appropriate solution.
Tianchi Lodge, a famous mountain hut in Taiwan, has operated an off-grid solar energy storage system with lithium iron phosphate (LFP) batteries since 2020.
In this case report, the energy architecture, detailed descriptions, and historical status of the system are provided.
\end{abstract}


\titlepgskip=-21pt

\maketitle

\section{Introduction}
\label{sec:introduction}

Mountain huts are buildings located at high altitude, in mountainous terrain, generally accessible only by foot, intended to provide food and shelter to mountaineers, hikers, and climbers~\cite{CanadianGeographies36(1992)144H.G.Kariel}.
They are known by many names, such as alpine hut, mountain shelter, mountain refuge, mountain lodge, mountain hostel, and mountain cabin.
Mountain huts can provide some functionalities and services.
The most important function is to provide shelter and simple sleeping berths for hikers to overcome the cold mountain environment at night.
Some mountain huts are not staffed, particularly in remote areas. However, others have staff who provide hot meals, drinks, and other services, such as activity spaces (for holding lectures, etc.), clothing sales, and small items.

Energy supply in high mountains remains an open issue to be solved.
Generally, grid connection is impractical, because the establishment and maintenance costs are too high.
It also causes many environmental problems, e.g., the destruction of the natural ecology in the mountain.
Diesel generators as energy suppliers have been used in high mountains for a long time despite their air pollution and noise problems.
Due to the development of renewable energy (such as solar, wind, and hydropower), the usage of diesel generators is reduced, lowering the emissions of greenhouse gases (GHGs). 
However, owing to their fluctuating nature, most renewable energy sources exhibit intermittent features. To deal with the problem of renewable energy intermittency, energy storage systems (ESSs) are necessary~\cite{IEEETrans.SmartGrid6(2015)124K.Rahbar, IEEETrans.Sustain.Energy1(2010)117S.Teleke}.
In the past, lead-acid batteries were heavily used as ESSs, accompanied by many environmental issues, e.g., poisoning, leaks, contamination of the environment, and damage to the ecosystem~\cite{Ecol.Indic.47(2014)210G.N.Liu, J.Hazard.Mater.250-251(2013)387X.F.Zhu}.
Recently, lithium-ion (Li-ion) batteries~\cite{EnergyStorageMater.1(2015)158J.B.Goodenough, Chem.Mater.22(2010)587J.B.Goodenough}, as a greener alternative, have started to replace lead-acid batteries in ESSs.

Tianchi Lodge, a famous ``five-star'' mountain hut in Taiwan~\cite{TianchiLodgeFiveStar(2023)Hikingnote, TianchiLodgeFiveStar(2023)bb705}, has adopted the off-grid solar energy storage system to provide electricity to the lodge since 2013.
However, lead-acid batteries were implemented as ESS, accompanied by many environmental issues.
In 2020, the lead-acid batteries were replaced by eco-friendly lithium iron phosphate (LFP) batteries.
A better and greener off-grid solar energy storage system has been established.

In this manuscript, a brief introduction of the Tianchi Lodge is given in Sec.~\ref{sec:Lodge}.
In Sec.~\ref{sec:ArchitectureEnergySystem}, the energy architecture of the off-grid solar energy storage system is presented with a detailed description of each component.
The historical development of energy systems is listed in Sec.~\ref{sec:HistoricalDevelopmentEnergySystems}.
Summary and Outlook are given at last (Sec.~\ref{sec:SummaryOutlook}).


\section{About Tianchi Lodge}
\label{sec:Lodge}

\subsection{Current status}

\begin{figure}[ht]
  \centering
  \includegraphics[width=\columnwidth]{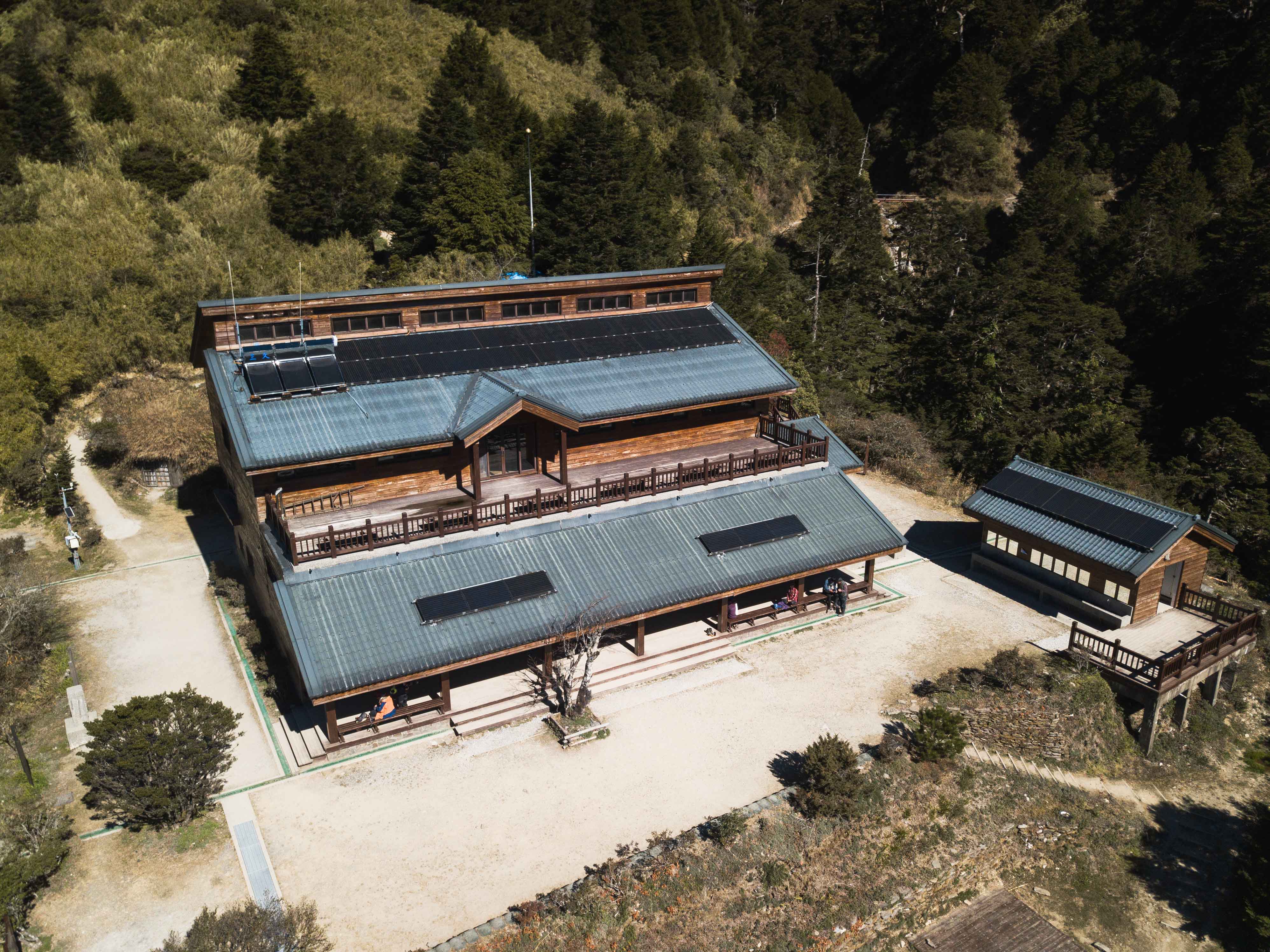}
  \caption{
  \justifying
  \textbf{Aerial Photography of Tianchi Lodge (2019).}
  (Photographer: Yu-Chun Lin)
  }
  \label{fig:TianchiLodgeAerialPhotography}
\end{figure}

Tianchi Lodge (coordinate: $24^\circ 02'43.15''$N, $121^\circ 16'46.77''$E) is an accommodation mountain lodge in Taiwan, located at an altitude of 2,860 m (9,380 ft)~\cite{TianchiLodgeStory()NantouFNCA}, at 13.1 km of the Nenggao Cross-Ridge Historic Trail (as shown in Fig.~\ref{fig:TianchiLodgeAerialPhotography}). It is situated in a sheltered mountain valley on the west side of the North Peak of Nenggao, under the jurisdiction of the Nantou Branch, Forestry and Nature Conservation Agency, Ministry of Agriculture, Taiwan. Its administrative area is located in Ren'ai Township, Nantou County.

The current Tianchi Lodge utilizes natural lighting, has solar panels installed on the roof, generates electricity using renewable energy, and is equipped with a lithium-ion battery ESS~\cite{TianchiLodgePost(2020)Wang-OurTrails}.
Its electricity can provide hot water and nighttime lighting for the lodge. The specially designed sloping roof allows rain and snow to flow down smoothly. The lodge is near two waterfalls, so there is no shortage of water.
The water is sufficient even during the dry season. It is an environmentally friendly green building.
The lodge also features a specially designed viewing platform on the second floor with large, operable glass windows, allowing visitors to enjoy the surrounding mountain scenery. The entrance corridor is spacious and includes a public social space with tea, tables, and chairs, providing hikers with a more comfortable activity space~\cite{TianchiLodgeReopen(2013)ETToday}. Due to these architectural designs and infrastructure, Tianchi Lodge has been dubbed a ``five-star'' luxury lodge in the mountain community by hikers~\cite{TianchiLodgeFiveStar(2023)Hikingnote, TianchiLodgeFiveStar(2023)bb705}.

Currently, Tianchi Lodge offers 3 rooms for twelve people, 4 rooms for eight people, and 5 rooms for four people, which can accommodate a total of 88 people~\cite{TianchiLodge-LodgeMap()TW-NantouFNCA}. As for the campsite, there are 26 campsites (23 small campsites and 3 large campsites). A total of 188 people can stay there each day~\cite{TianchiLodgeRooms(2018)Hikingnote,TianchiLodge-RoomApplication()TW-NantouFNCA}.

\subsection{Brief history}

Tianchi Lodge has been through several fires and reconstructions.
It was first built in 1918 and rebuilt three times in 1931, 1993, and 2012.

The Nenggao Garrison was the predecessor of the Tianchi Lodge. In September 1918, when the Nenggao Cross-Ridge Historic Trail~\cite{Book-StoryNengaoNationalTrail(2011)} was completed, the Nenggao Garrison was established for the needs of road maintenance and mail delivery~\cite{TianchiLodge01-TaiwanCulturalMemoryBank}.

In 1930, during the Musha Incident~\cite{MushaJiken-ITH-AcademicSinica}, the garrison was burned down. After the Musha Incident, the Japanese rebuilt the cypress house on the original site in 1931, with the same style as the Onoue Garrison (Yunhai Line Station)~\cite{TianchiLodge01-TaiwanCulturalMemoryBank}. Until it was burned down in 1986, it was the most famous high-altitude luxury accommodation among mountain climbers~\cite{TianchiLodgeStory()NantouFNCA,TianchiLodge01-TaiwanCulturalMemoryBank}

The Tianchi Lodge, rebuilt in 1993, was a tin-roofed mountain hut~\cite{TianchiLodgePost(2020)Wang-OurTrails} that could accommodate about 60 people. It was managed by the Forestry Bureau (now the Forestry and Nature Conservation Agency) and had a kitchen, a gas stove for cooking, and a diesel generator that provided electricity at night~\cite{Book-StoryNengaoNationalTrail(2011)}.

The Forestry Bureau commissioned ``Wu, Hsia-Hsiung Architect \& Associates'' in 2003 and 2005 to plan and design the renovation of Tianchi Lodge~\cite{Project-TianchiLodgeDesign(2003)TW-ChiayiFNCA, Project-TianchiLodgeDesign(2005)TW-FNCA}. Tianchi Lodge was demolished and rebuilt on its original site in 2011. The project was constructed by Sinyuan Construction Co., Ltd.~\cite{Project-TianchiLodgeNewBuild(2010)TW-NantouFNCA, Project-TianchiLodgeNewBuild-2ndChange(2012)TW-NantouFNCA} and supervised by Huangrueiming Architects~\cite{Project-TianchiLodgeSupervision(1998)TW-NantouFNCA, TianchiLodgeAward()TW-NantouFNCA}. The concept of green building design was introduced during the reconstruction. It is a two-story steel structure building with a roof facing due south. It is equipped with solar water heating panels (solar water heating system), a solar power generation system, and uses natural materials such as wood and gravel to replace premixed concrete. It is constructed with an earthquake-resistant steel structure and fireproof paint~\cite{TianchiLodgeAward()TW-NantouFNCA}. It has rooms, a restaurant, a kitchen, a toilet, and a solar power generation system~\cite{TianchiLodgeGreenBuildingOpen(2012)TW-LTN}.
It is planned to have five eight-person rooms, twelve three-person rooms, and four six-person rooms, which can accommodate one hundred mountain enthusiasts. It was completed in June 2012~\cite{TianchiLodgeGreenBuildingOpen(2012)TW-LTN} and opened to the public in 2013~\cite{TianchiLodgeReopen(2013)ETToday}. The Tianchi Lodge construction project won the ``2011-2012 Excellent Agricultural Construction Project Award (Architecture Category)''~\cite{TianchiLodgeAward()TW-NantouFNCA}.

Starting in May 2020, the Nantou Forest District Office, Forestry Bureau (now the Nantou Branch, Forestry and Nature Conservation Agency, Ministry of Agriculture) carried out a renovation project on the Tianchi Lodge for about four months, including exterior renovation, waterproofing, and surrounding facilities, in response to the policy of opening up the mountains and forests~\cite{Project-TianchiLodgeImprovement(2020)TW-NantouFNCA}.
The Tianchi Lodge is powered entirely by solar green energy. The lead-acid battery energy storage system used to store electricity was old and not environmentally friendly. During this renovation project, it was also replaced with a more efficient and environmentally friendly lithium-ion battery energy storage system, which can achieve better energy storage effect with a smaller volume~\cite{Project-TianchiLodgeGreenEnergy(2019)TW-NantouFNCA}.

\subsection{Alpine transportation supplement}

There are several main ways to transport mountain resources and equipment, such as manual transport, cable cars, and helicopters. The route from the Tunyuan trailhead in Ren'ai Township to the Nenggao Cross-Ridge Historic Trail leading to Tianchi Lodge is quite special, as it is one of the few routes in Taiwan where motorcycles and small four-wheeled transport vehicles can be used for transporting supplies.

However, due to the risk of accidents such as falling into valleys when using motor vehicles~\cite{MotorcycleDriverNenggaoOverpassDies(2020)TW-OurTrails}, the Nantou Forest District Office announced through the Tianchi Lodge booking website in August 2013 that, for safety reasons and to avoid accidents that could harm people, all personnel entering the Nenggao Cross-Ridge Historic Trail and Tianchi Lodge are strictly prohibited from using vehicles and bicycles except for official business that requires application and approval~\cite{MotorcycleDriverNenggaoOverpassDies(2020)TW-CNA}.

\section{Architecture of energy system}
\label{sec:ArchitectureEnergySystem}

Due to the difficulty of grid connection, energy supply in high mountains remains a challenging problem. Off-grid solar energy storage systems~\cite{Batteries10(2024)202H.C.Chung} provide a feasible solution for using renewable energy in high mountains and have been proven~\cite{Batteries10(2024)202H.C.Chung}.

At the end of 2019, the Nantou Forest District Office decided to install an off-grid solar energy storage system to supply energy to the lodge, and commissioned the C. J. Yan Architect \& Associate to carry out planning, design, and supervision~\cite{Project-TianchiLodgeSupervision(2019)TW-NantouFNCA}. The project was contracted by Lighting Art Co., Ltd.~\cite{Project-TianchiLodgeGreenEnergy(2019)TW-NantouFNCA}. Super Double Power Technology Co., Ltd. installed the system and carried out subsequent maintenance~\cite{Contract-TianchiLodgeSolarMaintenance(2024), Contract-TianchiLodgeSolarMaintenance(2023), Contract-TianchiLodgeSolarMaintenance(2022)}.

\begin{figure}[hb]
  \centering
  \includegraphics[width=\columnwidth]{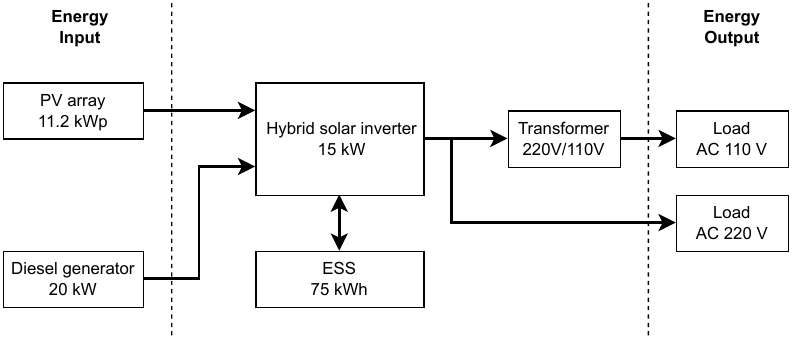}
  \caption{
  \justifying
  \textbf{Energy architecture of Tianchi Lodge (2020).}
  Energy input is provided in part by PV arrays and a diesel generator. The hybrid solar inverters directly supply power to the lodge's 220 V$_\mathrm{AC}$ load.
  The power of the 110 V$_\mathrm{AC}$ load is obtained through a transformer (220 V$_\mathrm{AC}$ / 110 V$_\mathrm{AC}$).
  An ESS stores and releases intermittent renewable energy from PV arrays. A diesel generator is used as the backup power source.
  }
  \label{fig:EnergyArchitecture}
\end{figure}

In 2020, the off-grid solar energy storage system of Tianchi Lodge was installed by Super Double Power Technology Co., Ltd., and began to supply energy to the lodge. Its energy architecture (as shown in Fig.~\ref{fig:EnergyArchitecture}) is generated by solar arrays, which charge the lithium-ion battery energy storage system~\cite{Sci.Data8(2021)165H.C.Chung, BookChChung2021EngIntePotentialAppOutlooksLiIonBatteryIndustry} through a hybrid solar inverter, and the inverter supplies power to the lodge's AC load. A diesel generator is used as the backup power source.

The details of the energy architecture are described below.

\subsection{Photovoltaic systems (PV systems)}

A photovoltaic (PV) system converts sunlight directly into direct current (DC) electricity using solar panels. Then the electricity is sent to a hybrid solar inverter for further power conversion.

In Tianchi Lodge, there are three PV arrays, providing a total power generation capacity of 11.2 kW$\mathrm{_p}$.
The first channel of the PV array is rooftop-mounted on the second floor of the lodge with a power generation capacity of 7.84 kW$\mathrm{_p}$.
The second channel of the PV array is rooftop-mounted on the first floor of the lodge with a power generation capacity of 1.68 kW$\mathrm{_p}$.
The third channel of the PV array is rooftop-mounted on the toilet with a power generation capacity of 1.68 kW$\mathrm{_p}$.
In order to integrate the PS arrays into the newly installed hybrid solar inverter system, the series and parallel connections of the PV arrays must be readjusted to meet the inverter input specifications.


\subsection{Hybrid solar inverters}

The hybrid solar inverter serves the crucial role of linking PV systems, ESS, AC loads, and backup power sources.
The total nominal power of the hybrid solar inverter system is 15 kW.
Considering the challenges of maintaining mountain huts, a system design employing multiple units operating in parallel to simultaneously output power was chosen for reliability and stability. Under normal operating conditions, all units can supply power to the lodge. In the event of any failure, the faulty unit can be disconnected, and the remaining working units can continue supplying power while the system provider is notified for maintenance. During this process, the inverter units can still supply power to the lodge, significantly reducing downtime caused by system failures.

Some features of the hybrid solar inverters are listed:
(1) detachable LCD control Panel,
(2) built-in Wi-Fi function for APP monitoring,
(3) zero (0 ms) transfer time, the best protection for servers,
(4) pure sine wave inverter with built-in MPPT solar charger,
(5) selectable power charging current,
(6) configurable AC/Solar input priority via LCD setting,
(7) auto-restart while AC is recovering,
(8) overload and short circuit protection,
(9) cold start function, and
(10) optional parallel operation of up to 9 pcs.

\subsection{LFP energy storage system (ESS)}

ESS is a technology that stores excess electricity and releases it when needed, acting as a ``reservoir'' in the power system. It is mainly used to stabilize the power grid, integrate intermittent renewable energy sources (such as solar, wind, and hydropower), and provide peak shaving and valley filling to improve power supply efficiency and reliability. Its core components include battery modules and BMS, and it regulates power supply and demand through charging and discharging.

ESSs in high-altitude areas have long used lead-acid batteries. However, the heavy metal lead and highly acidic electrolyte contained in these batteries can severely pollute soil and water sources and harm human health (nervous system, kidneys, and brain) if discarded or improperly disposed of. Without adequate equipment, the recycling and processing process can also generate pollutants such as lead dust and acid mist.

To enable the use of green energy in high-altitude areas, we need the support of new technologies. The vigorous development of electric vehicles (EVs) in recent years has also driven the progress of Li-ion battery technology. Compared with lead-acid batteries, which have a history of over a century but are highly polluting, Li-ion batteries have many advantages, such as high energy density, light weight, no memory effect, low self-discharge rate, and long lifespan. Li-ion batteries have been widely used in 3C products and EVs. The 2019 Nobel Prize in Chemistry was awarded to John B. Goodenough, M. Stanley Whittingham, and Akira Yoshino for their contributions to the development of Li-ion batteries ~\cite{NobelPrizeChem2019NPO, NobelPrizeChem20191009CNN, NobelPrizeChem20191009Guardian}.

Through long-term technological improvements, lithium-ion batteries have evolved into many different types (such as ternary Li-ion batteries and LFP batteries), each with its own advantages and disadvantages. For example, ternary Li-ion batteries have high energy density but slightly weaker safety, while LFP batteries, although having slightly lower energy density, are safe, stable, and have a long cycle life. To address the challenges of maintenance in high-altitude environments, safe and highly stable LFP batteries are chosen as the energy storage system.

The LFP batteries~\cite{Sci.Data8(2021)165H.C.Chung} are used with BMS in the energy storage system. The batteries are in series connection (16S configuration), making a battery system with a nominal voltage of 48 V$_\mathrm{DC}$ and a capacity of about 75 kWh. A balance circuit is also applied in the system for balancing the voltage of each battery cell. The information of the ESS can be sent to the EMS via the RS-485 communication port.

\subsection{Battery management system (BMS)}

To use the battery safely, a well-designed battery management system (BMS) is required. A BMS is any electronic system that manages a battery cell, pack, or module system, such as ensuring the battery operating within its safe operating area, monitoring its status, calculating secondary data, reporting that data, controlling its environment, and cell balancing~\cite{BookChChung2021EngIntePotentialAppOutlooksLiIonBatteryIndustry}.

\subsection{Loads}

Two AC load channels with different voltages (220 V$_\mathrm{AC}$ and 110 V$_\mathrm{AC}$) and the same frequency of 60 Hz are used in the lodge.
When designing the system, it is necessary to assess the power consumption of the lodge during normal operation and during high-load operation.

\subsection{Backup power system}

A diesel generator is used as the backup power system.
When the power from the PV system is insufficient (e.g., rainy days or cloudy days), the diesel generator can supply the insufficient power to the lodge.
When the system encounters failures (e.g., temporary shutdown of the hybrid solar inverter), the diesel generator can supply the full power to the lodge.

In Tianchi Lodge, a three-phase diesel generator with nominal output voltage 220 V$_\mathrm{AC}$ and frequency of 60 Hz is installed on site.
The output power capacity is 20 kW.

\subsection{Energy management system (EMS)}

No energy management system (EMS) is installed in this system.
The energy information, such as energy generation, usage, and storage information, is not available.
It's difficult to realize the energy input and output of the system quantitatively.



\begin{figure}[ht]
  \centering
  \includegraphics[width=\columnwidth]{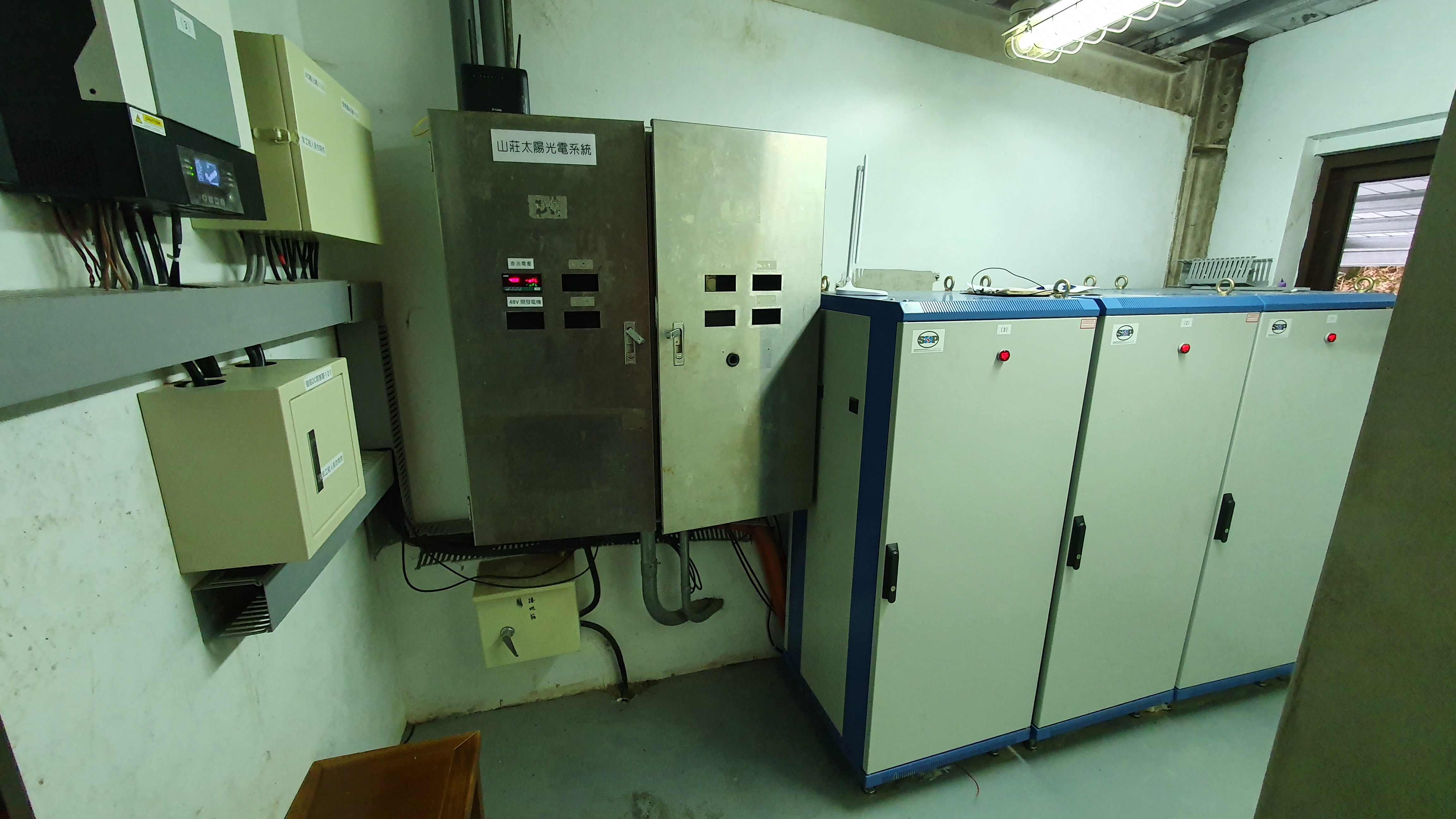}
  \caption{
  \justifying
  \textbf{Machine room of Tianchi Lodge (2022).}
  }
  \label{fig:MachineRoom}
\end{figure}

\section{Historical development of energy systems}
\label{sec:HistoricalDevelopmentEnergySystems}

\begin{table*}[htb]
  \centering
  \begin{tabular}{| c | c | c | c | c | c |}
    \hline
    \textbf{Year} & \textbf{Renewable energy} & \textbf{Energy storage system} & \textbf{Load capacity} & \textbf{Energy management system} & \textbf{Project} \\
    \hline
    2020 & Total: 11.2 kW$\mathrm{_p}$ (PV) & Lodge: 75 kWh (LFP) & 15 kW & No & \cite{Project-TianchiLodgeGreenEnergy(2019)TW-NantouFNCA} \\
    \hline
    2013 & Total: 11.2 kW$\mathrm{_p}$ (PV) & Lodge: 86 kWh (Lead-acid) & 5 kW & No & \cite{Project-TianchiLodgeNewBuild(2010)TW-NantouFNCA} \\
     &                      & Toilet: 9.6 kWh (Lead-acid) &  &  &  \\
     & Component:                            &  &  &  &  \\
     & Lodge (2F): 7.84 kW$\mathrm{_p}$ (PV) &  &  &  &  \\
     & Lodge (1F): 1.68 kW$\mathrm{_p}$ (PV) &  &  &  &  \\
     & Toilet: 1.68 kW$\mathrm{_p}$ (PV)     &  &  &  &  \\
    \hline
  \end{tabular}
  \caption{Evolution of the Energy System of Tianchi Lodge}
  \label{tab:EnergySystemEvolution}
\end{table*}

In the past, lead-acid battery ESSs combined with diesel generators were commonly used for the power supply of mountain huts. However, the environmental damage caused by lead pollution from lead-acid batteries and their harm to human health have created serious problems for their use in high-altitude areas. Furthermore, lead-acid batteries will age and deteriorate after a few years of use, eventually becoming unable to store electricity.

When Tianchi Lodge officially opened in 2013~\cite{TianchiLodgeReopen(2013)ETToday}, it introduced facilities such as a solar power generation system~\cite{TianchiLodgeGreenBuildingOpen(2012)TW-LTN}. However, the ESS at that time was built using lead-acid batteries, which are less environmentally friendly.

In terms of PV arrays, 1.68 kW$_\mathrm{p}$ solar panels were installed on the roof of the first floor of the lodge, 7.84 kW$_\mathrm{p}$ solar panels were installed on the roof of the second floor of the lodge, and 1.68 kW$_\mathrm{p}$ solar panels were installed on the roof of the toilet. In terms of energy storage, the lodge had 86 kWh of lead-acid batteries, and the toilet had 9.6 kWh of lead-acid batteries. The inverter had an AC output capacity of 5 kW and a rated voltage of 220 V$_\mathrm{AC}$. The diesel generator had a power generation capacity of 20 kW and a rated voltage of 220 V$_\mathrm{AC}$.

The evolution of the energy system in Tianchi Lodge is shown in Table~\ref{tab:EnergySystemEvolution}.

%
%

\section{Summary and Outlook}
\label{sec:SummaryOutlook}

Using renewable energy and eco-friendly ESS in high mountains is not easy.
It took eight years to reach the goal in Tianchi Lodge.
Since 2013, renewable energy (PV systems) has been used.
However, the lead-acid battery ESS still causes pollution in high mountains.
Until 2020, the eco-friendly LFP ESS replaces the retired lead-acid battery ESS, a better and green off-grid solar energy storage system is built.
After years of use, its operational reliability has been verified.

In the next step, we should install an EMS to check the energy input and output.
Figuring out the energy usage of the lodge.
Further suggestions can be given after the energy usage analysis, such as
``Is the solar power generation sufficient, and is it necessary to expand the solar power system?'' or ``How many days can the ESS withstand without solar power generation, and is it necessary to expand the ESS?'' Although diesel generators serve as a backup solution for the backup energy system, we still hope that their usage time can be as short as possible to minimize their damage to the high-altitude environment.

We wish not only for a green building in high mountains, but also for using green energy to protect the ecosystem and environment.

\section*{Acknowledgment}
The author (H.-C. Chung) would like to thank the contributors to this article for their valuable discussions and recommendations: Jung-Feng Jack Lin, Hsiao-Wen Yang, Yen-Kai Lo, and An-De Andrew Chung.
The author (H.-C. Chung) thanks Pei-Ju Chien for English discussions and corrections, as well as Ming-Hui Chung, Su-Ming Chen, Lien-Kuei Chien, and Mi-Lee Kao for financial support.
This work was supported in part by Super Double Power Technology Co., Ltd., Taiwan, under the project ``Development of Cloud-native Energy Management Systems for Medium-scale Energy Storage Systems (\href{https://osf.io/7fr9z/}{https://osf.io/7fr9z/})'' (Grant number: SDP-RD-PROJ-001-2020).


%
%

\begin{CJK}{UTF8}{bsmi}


\end{CJK}

\begin{IEEEbiography}
[{\includegraphics[width=1in,height=1.25in,clip,keepaspectratio]{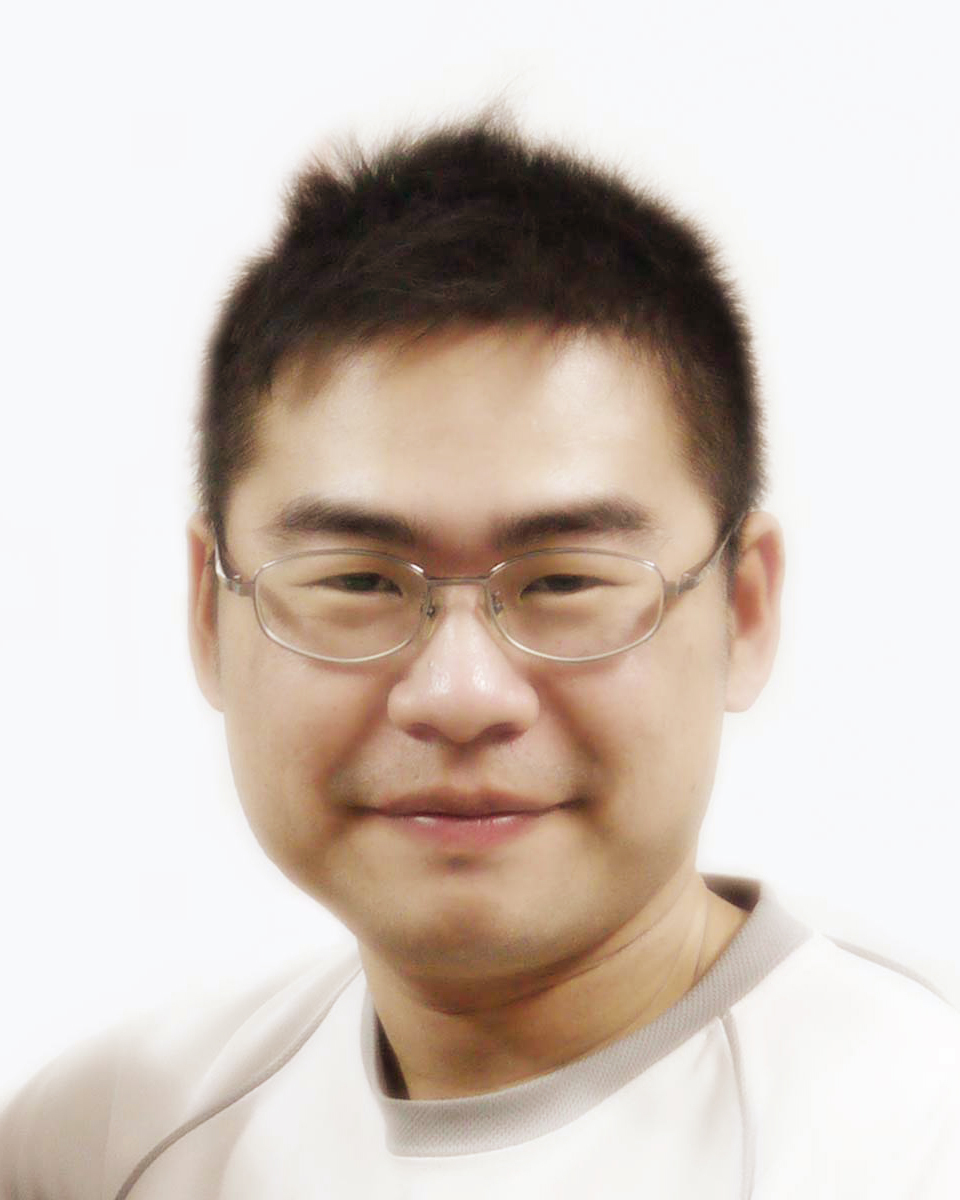}}]
{Hsien-Ching Chung} (M'17--SM'22) received the B.S., M.S., and Ph.D.  degrees in physics from the National Cheng Kung University, Tainan, Taiwan, in 2011.

Dr. Chung has complete experience in the lithium-ion (Li-ion) battery industry supply chain, from the battery cell factory, pack factory, to the system integration factory. Recent interests focus on system integration and applications of Li-ion battery energy storage systems. Currently, he is conducting the RD project "Development of Cloud-native Energy Management Systems (EMSs) for Medium-scale Energy Storage Systems." The cloud-native EMS has been installed at the Paiyun lodge (the highest lodge) in Taiwan.

He also has rich experience in fundamental research. From 2011 to 2017, as a postdoctoral fellow, his main scientific interests in condensed matter physics included the electronic and optical properties of carbon-related materials, low-dimensional systems, and next-generation energy materials.

Dr. Chung is a senior member of the Institute of Electrical and Electronics Engineers (IEEE), a member of the American Physical Society (APS), a member of the American Chemical Society (ACS), and an associate member of the Institute of Physics (IOP).
\end{IEEEbiography}

\EOD

\end{document}